\documentclass[1p,preprint]{elsarticle}

\usepackage[utf8]{inputenc}
\usepackage{amsmath,graphicx}
\usepackage{amsfonts}
\usepackage{amssymb}
\usepackage{color}
\begin{document}
\begin{frontmatter}

\title{Phase-locking dynamics of heterogeneous oscillator arrays}
\author[lepri1,mysecondaryaddress]{Stefano Lepri}
\address[lepri1]{Consiglio Nazionale delle Ricerche, Istituto dei Sistemi Complessi, Via Madonna del Piano 10 I-50019 Sesto Fiorentino, Italy} 
\address[mysecondaryaddress]{Istituto Nazionale di Fisica Nucleare, Sezione di Firenze, via G. Sansone 1 I-50019, Sesto Fiorentino, Italy}
\author[ap,ap2]{Arkady Pikovsky}
\address[ap]{Department of Physics and Astronomy, University of Potsdam
Karl-Liebknecht-Str 24/25, Bldg 28    D-14476, Potsdam, Germany}
\address[ap2]{Department of Control Theory, Lobachevsky University of Nizhny Novgorod, Gagarin Avenue 23, 603950 Nizhny Novgorod,
Russia}

\begin{abstract}
We consider an array of nearest-neighbor coupled nonlinear autonomous oscillators 
with quenched random frequencies and purely conservative coupling. We show that global phase-locked states emerge in finite lattices and study numerically
their destruction. Upon change of model parameters, such states are found to become unstable with the generation of localized periodic and chaotic oscillations. 
For weak nonlinear frequency dispersion,  metastability occur
akin to the case of almost-conservative systems. 
We also compare the results with the phase-approximation in
which the amplitude dynamics is adiabatically eliminated.    
\end{abstract}
\end{frontmatter}

\section{Introduction}

Nonlinear dynamics of  
systems  driven out of equilibrium by external 
(possibly non-conservative) forces is a fascinating research
subject. The aim of this vast, interdisciplinary research is to illustrate 
generic and universal features by simple paradigmatic models. 
Among them, systems of 
\textit{classical coupled oscillators} are of particular interest as they represent
a large variety of different physical problems like atomic vibrations
in crystals and molecules or field modes in optics or acoustics.  
 
Among a large number of nonlinear effects, synchronization 
of coupled autonomous oscillators, first empirically observed by Huygens, is one of the most remarkable.
The basic theory has been advanced considerably 
\cite{pikovsky2003synchronization} but still attracts
considerable attention, as demonstrated by the huge number of papers
published in the last decade.

A  notable source of complexity in nonlinear systems is heterogeneity of individual elements.
For ensembles of linear, heterogeneous oscillators, static disorder is known
to lead to Anderson localization of matter, light, and
sound waves. Generically, the normal models 
(eigenstates) display a localized pattern that usually hinders energy 
propagation \cite{Sheng2006}.  

As a matter of fact, in presence of disorder, nonlinearity, and interactions  a whole new set
of phenomena may arise. 
The interplay between disorder and nonlinearity, 
their localizing and delocalizing effects on lattice waves is also 
 an intriguing and challenging current research theme. 
For conservative systems, it has been recognized that the addition of nonlinearity 
causes interaction among the eigenmodes, which results
in a slow wave diffusion \cite{Pikovsky08,Skokos2009,mulansky2011scaling}. 
For instance, two different regimes of destruction of Anderson localization 
(asymptotic weak chaos, and intermediate strong chaos), separated by a crossover condition on densities have been reported \cite{Flach2016}. 
External driving and dissipation can lead to even richer scenarios.
For instance, disordered nonlinear array with forcing and dissipation admit  Anderson attractors \cite{laptyeva2015anderson} -- stationary multipeak patterns composed by a set of interacting, 
excited Anderson modes. Such attractors emerge by  
joint effect of the pumping-induced mode excitation, nonlinearity-induced mode 
interactions, and stabilization by dissipation. 

Besides intrinsic hetereogeneity of oscillators, another source 
of disorder may arise from topology of connections, as it
occurs for coupled nonlinear oscillator on networks (see e.g. 
ref.\cite{dipatti2018ginzburg} and references therein).
From the experimental point of view, we mention active optical systems,
where joint effect of disorder (both in frequencies and 
cavity topology) and nonlinearity is relevant for random lasers
\cite{conti2011complexity}
and lasing networks \cite{lepri2017,lepri2020chaotic}.

If coupling of periodic self-sustained oscillators is weak, the dynamics can be described
in terms of the phase approximation, where only a variation of
oscillator phases are considered. For two coupled oscillators
this leads to an Adler-type equation \cite{pikovsky2003synchronization}. The corresponding phase models are extensively used to study oscillator
lattices and globally coupled ensembles \cite{kuramoto2012chemical}.
Much less studied is the case in which the amplitude dynamics enters
into play. 

A related important issue is the possibility of observing states with a non-trivial
collective dynamics. In the context of coupled phase oscillators, the
case of short-ranged coupling (e.g. to nearest-neighbor on a lattice)
has been less considered than the globally coupled one. 
A general question regards the distinction between
synchronization and frequency entrainment and the mapping 
to the diffusion equation \cite{hong2005collective}. 
The linear theory predicts frequency entrainment in 
any dimension, for any non zero coupling but the role of
nonlinear terms is still to be understood.
It is known \cite{ermentrout1984frequency} that 
up to certain values of disorder in natural
frequencies of phase oscillators, one can find synchronous
states. Also, it has been proved that the probability of frequency entrainment 
is generically exponentially small in the number of 
oscillators \cite{strogatz1988phase}. 

In the present work we report a numerical study of a  one-dimensional
lattice of self-sustained heterogeneous oscillators described by their 
phase and amplitudes with purely reactive (i.e.\textit{conservative}) coupling. 
More precisely, we consider a one-dimensional 
Ginzburg-Landau lattice \cite{aranson1986strange} where each site 
has a different (random) natural frequency.
In the literature,  the case of purely 
diffusive coupling and global coupling it is mostly considered, see e.g.   
\cite{blasius2003anomalous,yang2007transitions,daido2011dynamics}.
An example of global reactive coupling is in Ref. \cite{cross2006synchronization}.
The choice of a purely conservative and local coupling 
is motivated by the experiments on lasing networks, where 
the individual elements (non-identical active and passive 
fibers) are coupled by non-dissipative optical couplers
\cite{giacomelli2019optical} but applies to more general 
nonlinear multimode photonic networks \cite{ramos2020optical}.
A similar coupling has been recently realized for nanoelectromechanical 
oscillators ~\cite{Matheny_etal-19}.
We will illustrate that non-trivial state exist, including frequency-entrained domains and localized chaotic states. The paper is organized as follows.
In Section \ref{sec:model} we define the model and in Section \ref{sec:simul} we 
describe some of its phenomenology. Section \ref{sec:ens} deals with 
average properties with respect to the disordered realizations. 
A brief discussion on the phase approximation and its continuum limit
is given in Section \ref{sec:phase}. 

\section{Model}
\label{sec:model}

In the present work we considered a specific paradigmatic model
namely the \textit{discrete Ginzburg-Landau} 
(or locally coupled Stuart-Landau) array of oscillators  with with \textit{random intrinsic frequencies}
\begin{equation}
\dot{\psi}_n  =  \left[i\omega_n+ \gamma + (i\alpha - \beta)  |\psi_n|^2\right]  \psi_n 
+ (c_2+ic_3)(\psi_{n-1}+\psi_{n+1})
\label{cgl}
\end{equation}
where $\psi_n$ is the complex oscillation amplitude. Frequencies $\omega_n$ are chosen as
independent random numbers
uniformly distributed in $[-w/2,w/2]$, they produce \textit{quenched disorder}. 
The index $n=1,\ldots,N$ labels the lattice sites and 
periodic boundary conditions are assumed. The parameter $\alpha$
describes nonlinear frequency shift while $\beta>0$ is the nonlinear
dissipation which is necessary to counterbalance the growth term 
$\gamma>0$. 

We focus on the \textit{purely reactive coupling} case 
namely we set $c_2=0$ and $c_3\equiv c\neq 0$. The model has two interesting limits.
For vanishing $\alpha,\beta,\gamma$ it reduced to
the well-known one-dimensional Anderson model (linear disordered lattice). For $\gamma=\beta=0$ 
and $\omega_n$ non-random, it
is the so-called Discrete Nonlinear Schr\"odinger equation \cite{Kevrekidis}, which 
has been widely investigated in many different
nonequilibrium conditions \cite{franzosi2011discrete,borlenghi2014thermomagnonic,iubini2017chain}.
The DNSE with random frequencies has been considered in \cite{Pikovsky08,Skokos2009}.

For later reference, we also write the equations for amplitude and phases 
of the complex field ${\psi}_n=a_n e^{i\theta_n}$.
Introducing the phase differences $\varphi_n=\theta_{n+1}-\theta_n$ 
and the relative detunings  $\delta_n=\omega_{n+1}-\omega_n$
between neighbouring oscillators, we obtain the exact equations
\begin{eqnarray}
\label{ampphas}
&& \dot a_n = (\gamma - \beta a_n^2 ) a_n +c \left[a_{n+1}\sin \varphi_n -
a_{n-1}\sin \varphi_{n-1}  \right] \nonumber\\
&& \dot\varphi_n= \delta_n  +\alpha (a_{n+1}^2-a_{n}^2)+ \\
&&+c\left[
\frac{a_{n+2}}{a_{n+1}}\cos \varphi_{n+1} +
\left(\frac{a_{n}}{a_{n+1}}- \frac{a_{n+1}}{a_n}  \right)\cos \varphi_{n} -
\frac{a_{n-1}}{a_n}\cos  \varphi_{n-1} \right]
\nonumber
\end{eqnarray}
Note that the $\delta_n$ are random (with uniform distribution) 
but for periodic boundary 
they must satisfy the constraint $\sum_n \delta_n=0$ (only the first $N-1$ differences 
are indeed independent).
Correspondingly, the number of equations for the phase differences $\varphi_n$ is $N-1$.
Those equation will be the basis of some approximate treatment
below.

The model has five control parameters but, upon rescaling, one may fix two to unity. In the following, we choose to leave the gain parameters
$\gamma$ free, and we set $c=1$, $\beta=1$
so that $w$ and $\gamma$ are in units of $c$.

\section{Simulations}
\label{sec:simul}

We have first performed a series numerical simulations
for different gain parameters $\gamma$ and $N=200$, and fixed $w,\alpha$. 
Several lattice sizes have been considered with similar 
outcomes. 
The monitored quantities are the time averages (over  the simulation time $T$)
of the amplitudes
$\langle |\psi_n|^2 \rangle$ and the mean frequencies
\[
\langle\dot{\theta_n}\rangle = \frac{1}{T} [\theta_n(T)-\theta_n(0)]\;,
\]
as well as fluctuations of these quantities 
$\langle \dot\theta_n ^2 \rangle - \langle \dot\theta_n  \rangle^2$ and $\langle |\psi_n|^4 \rangle - \langle |\psi_n|^2 \rangle^2$,
respectively.

\begin{figure}
\begin{center}
\includegraphics[width=\textwidth]{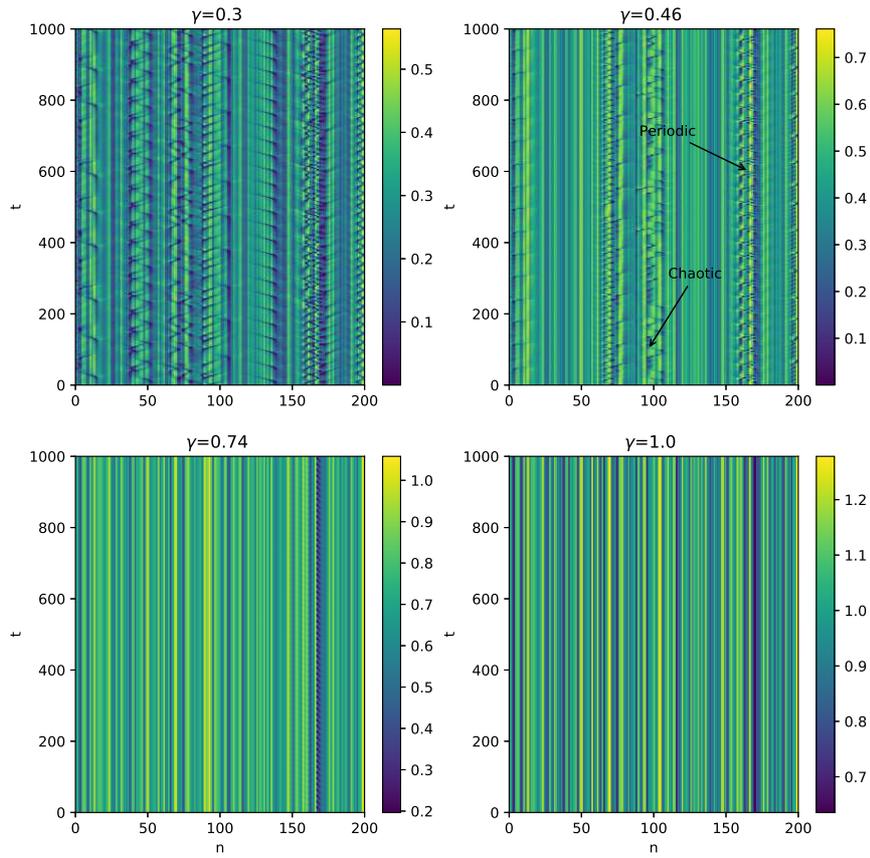}
\end{center}
\caption{Space-time plots of $|\psi_n|^2$ for different  $\gamma$ and $w=1.5$ 
$\alpha=2$, chain
length $N=200$, after discarding a transient of $10^4$ time units.
At large $\gamma=1$ a stationary state is observed. At $\gamma=0.74$,
periodic modulation close to $n=170$ occurs. At $\gamma=0.46$, several domains with perdiodic and 
chaotic modulations are observed. This state can be described as 
a coexistence of phase-locked clusters separated by localized chaotic
or periodic oscillations. Finally, at $\gamma=0.3$, the regime is rather turbulent.}
\label{fig:figure1}
\end{figure}

\begin{figure}
\begin{center}
\includegraphics[width=\textwidth]{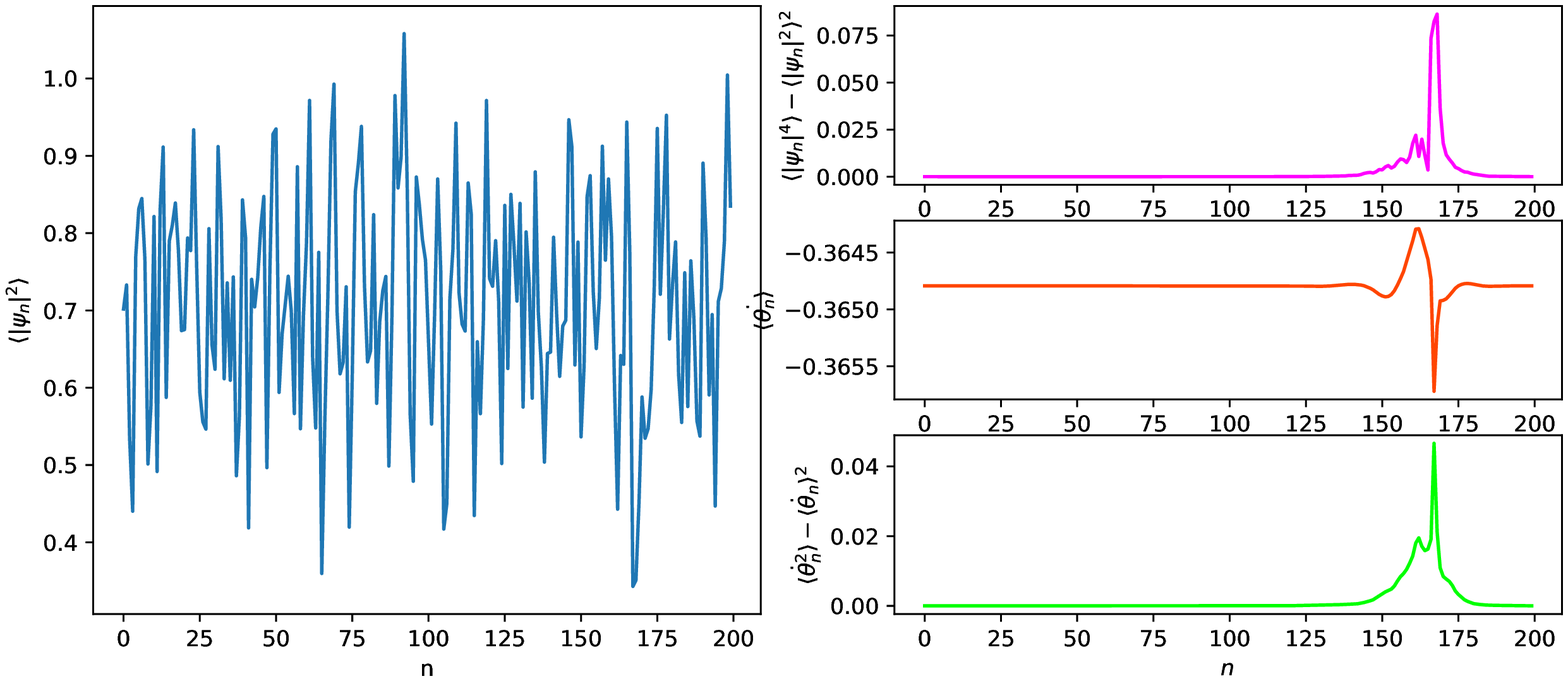}
\includegraphics[width=\textwidth]{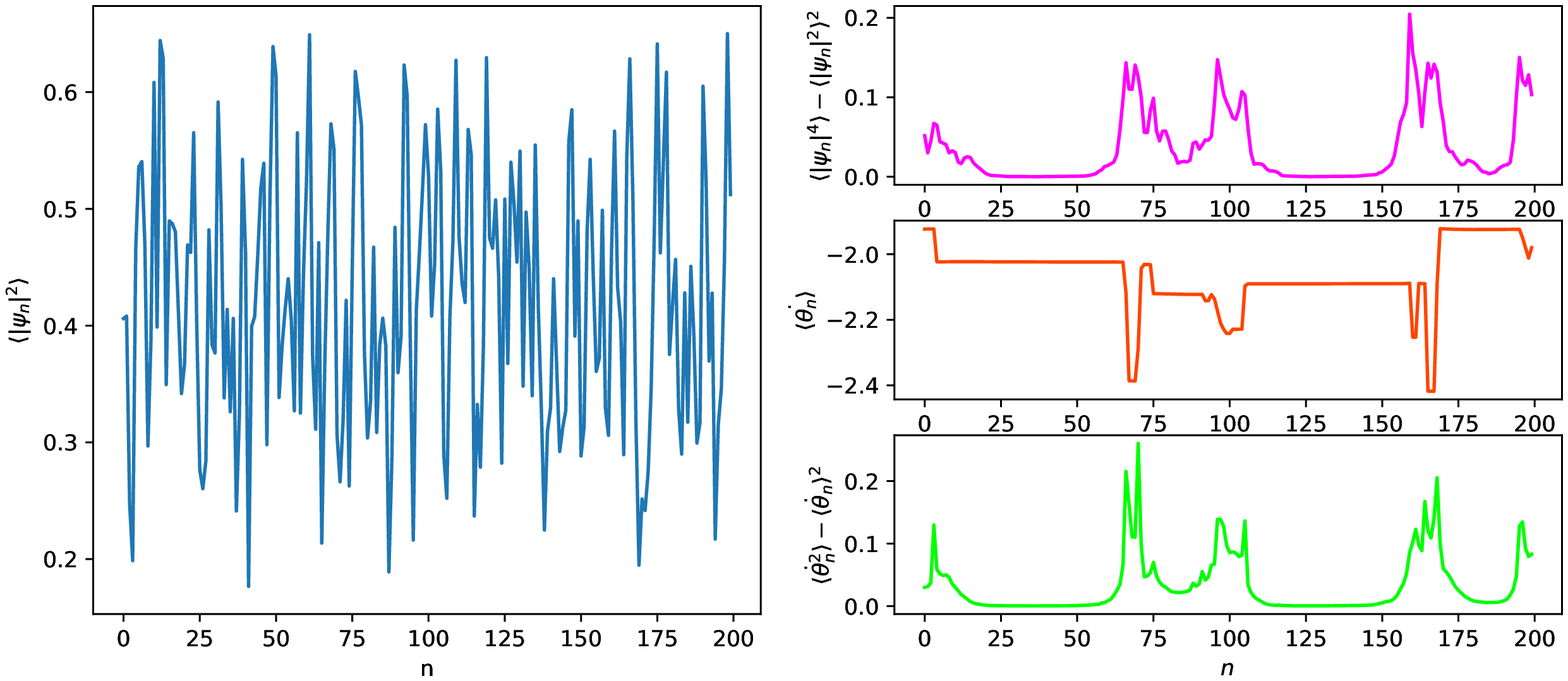}
\end{center}
\caption{Spatial configurations for $\gamma=0.74$ (upper panel)
and$\gamma=0.46$ (lower panel) of the simulations presented
in figure~\ref{fig:figure1}.
In each panel the average amplitudes and their 
fluctuations, the average frequencies and their fluctuations. 
For $\gamma=0.74$ there is only one irregular region; for $\gamma=0.46$ there are four 
such domains.
}
\label{fig:figure2}
\end{figure}

\begin{figure}
\begin{center}
\includegraphics[width=0.45\textwidth]{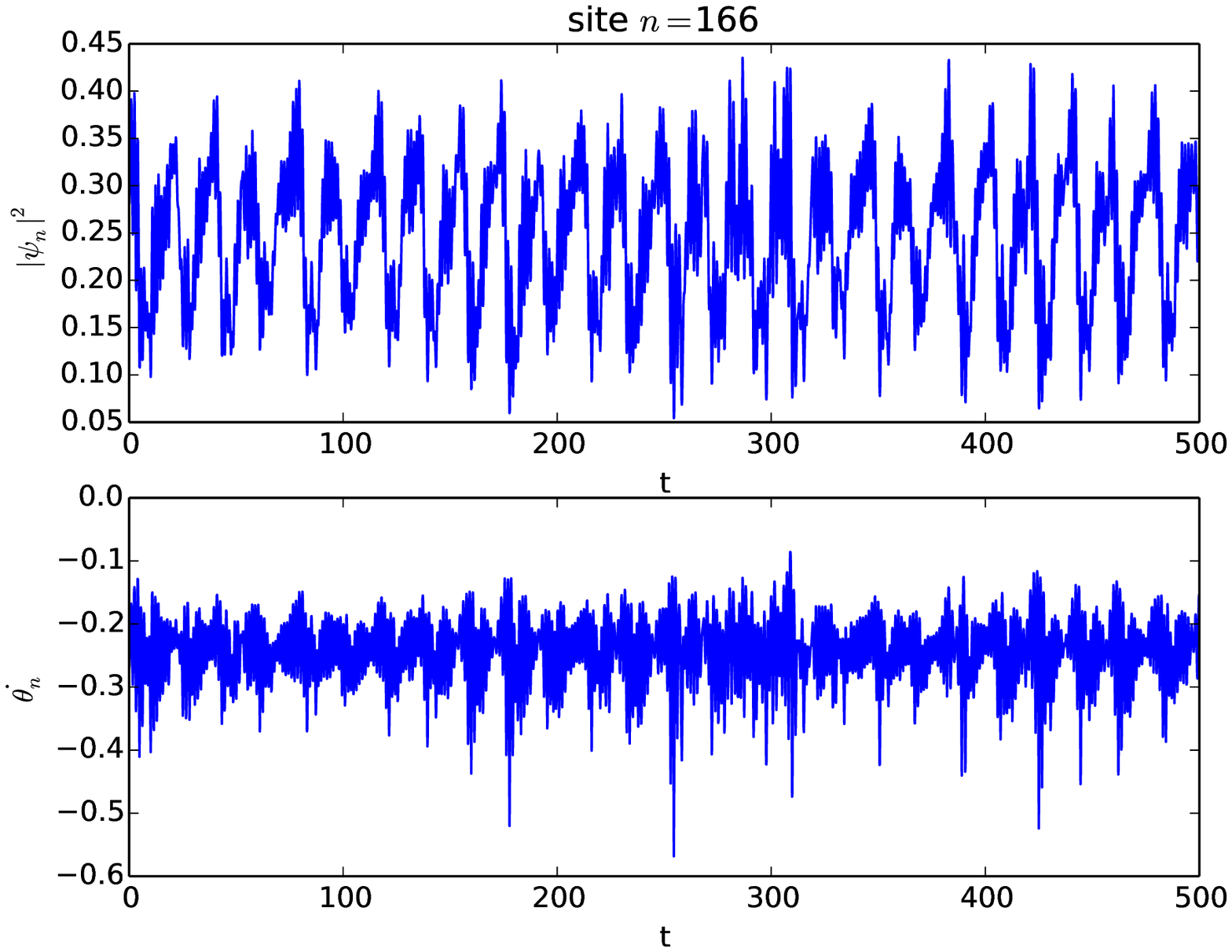}
\includegraphics[width=0.45\textwidth]{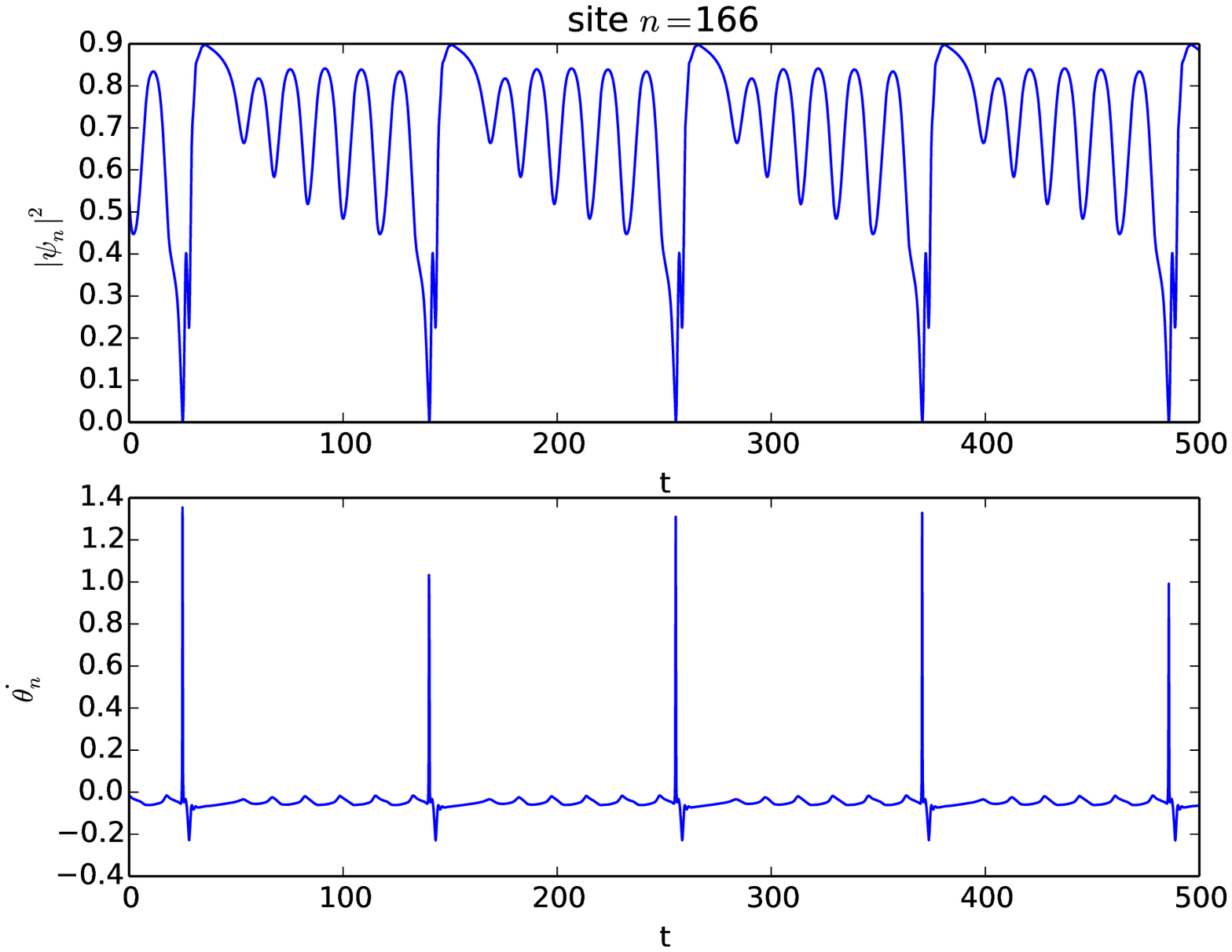}
\includegraphics[width=0.45\textwidth]{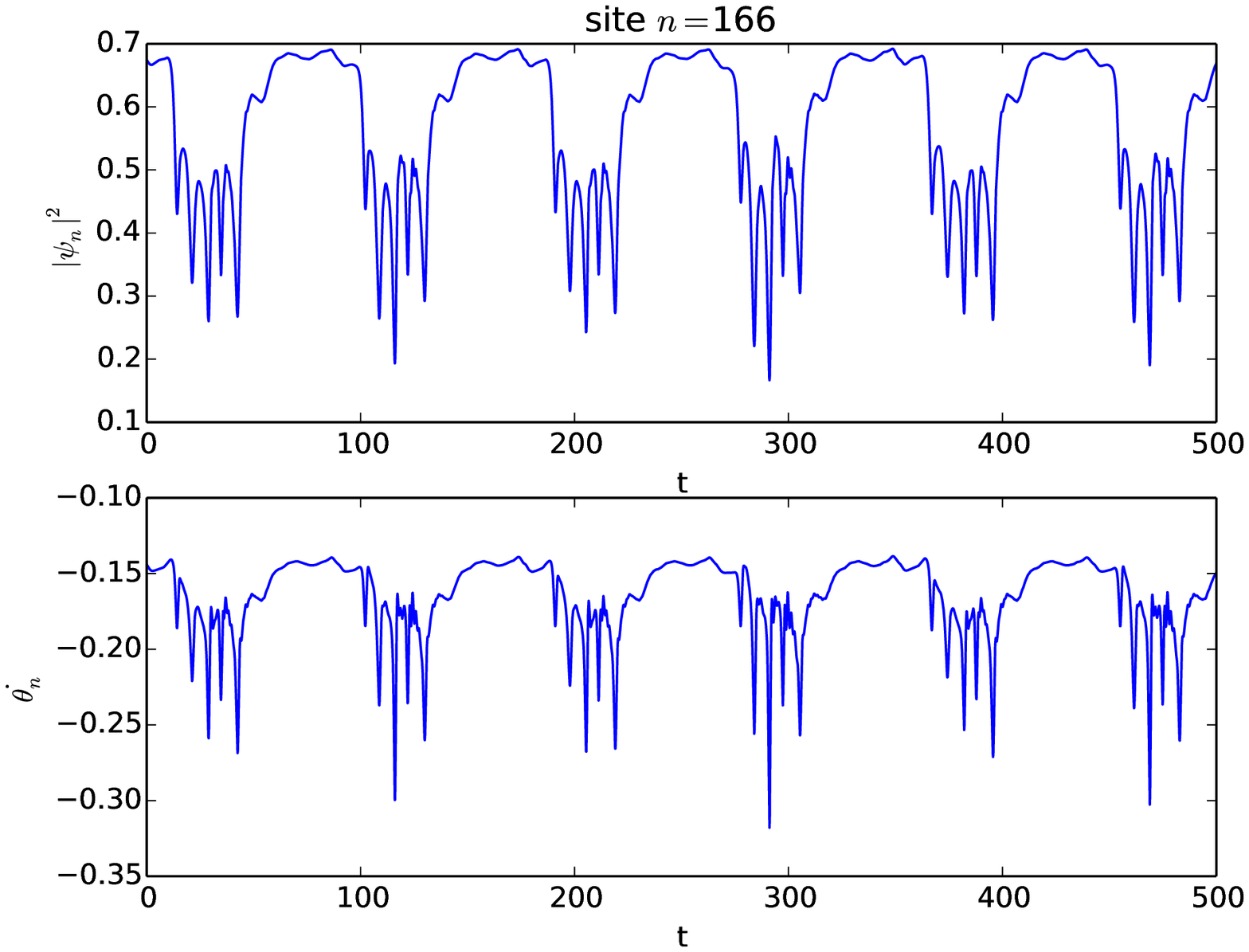}
\includegraphics[width=0.45\textwidth]{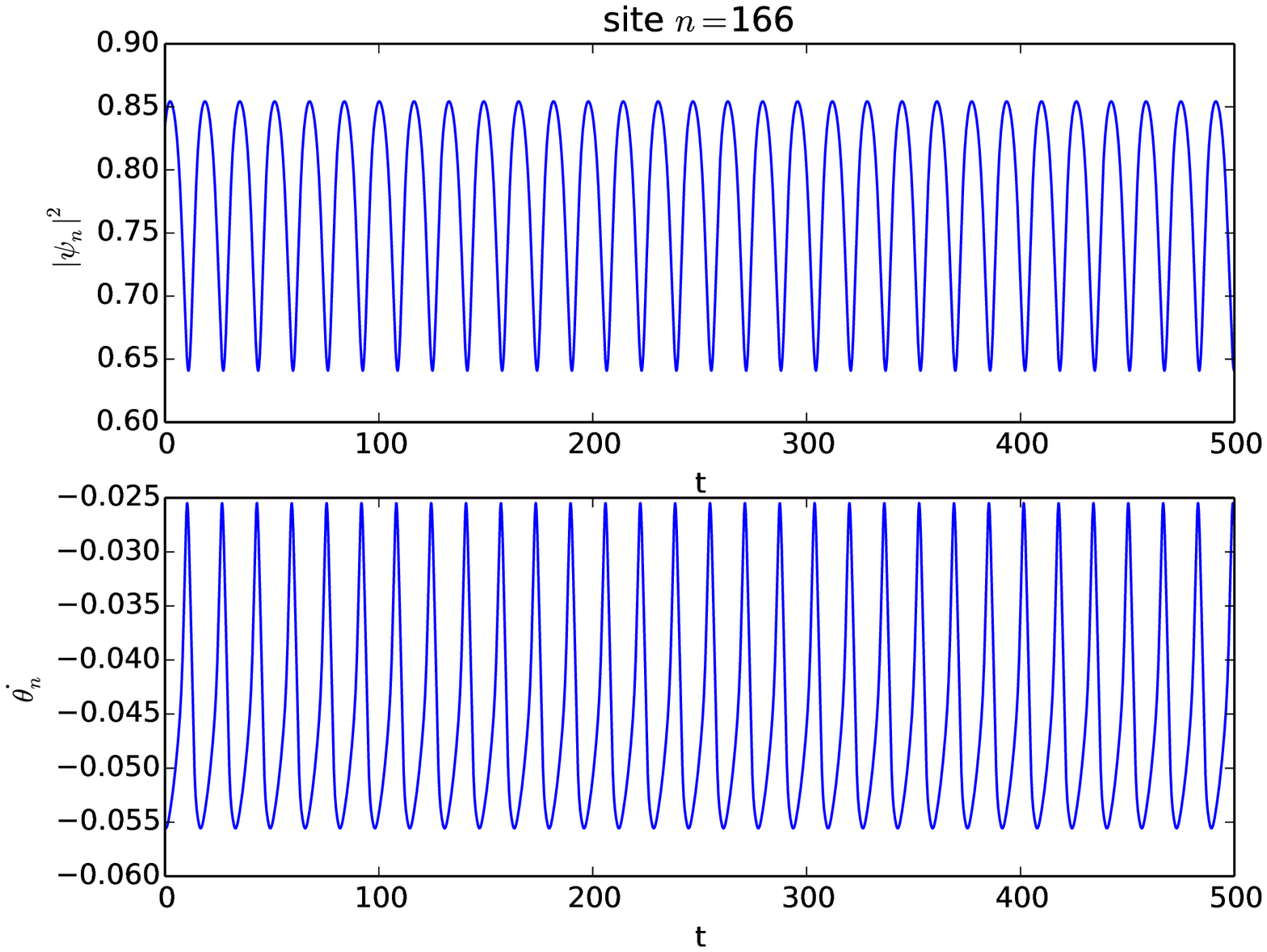}
\end{center}
\caption{Time evolution of amplitudes and phase velocities at the site 
$n=166$ of the chain for the simulation reported above. 
Upon decreasing $\gamma$ there is a bifurcation from 
periodic to chaotic. Parameters are the same as in fig.\ref{fig:figure1}.
}
\label{fig:figure3}
\end{figure}


\begin{figure}
\begin{center}
\includegraphics[width=\textwidth]{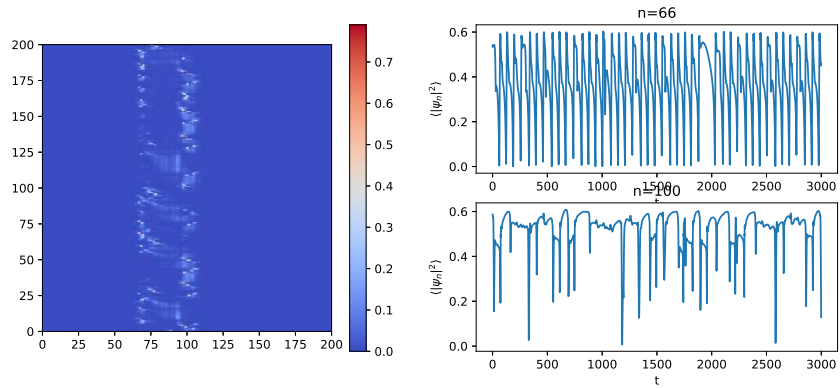}
\end{center}
\caption{Localized chaos for $\alpha=2$ and $w=1.5$, $\gamma=0.46$, chain
length $N=200$, calculated after discarding a transient of $10^4$ time units.
Left: Space-time plot of the Lyapunov vector displaying
a localized structure around site $n=100$. Right: time traces of the 
oscillator amplitudes on two sites belonging to the chaotic region.
}
\label{fig:figure5}
\end{figure}

\begin{figure}
\begin{center}
\includegraphics[width=\textwidth]{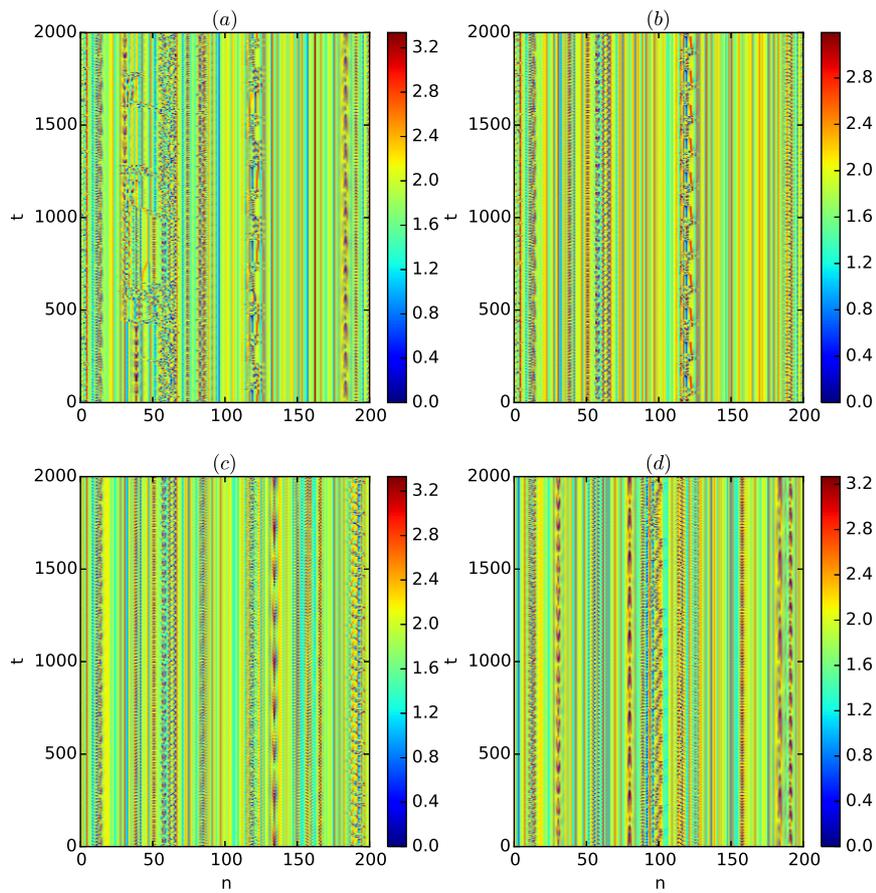}
\end{center}
\caption{Space -time plots of $|\psi_n|^2$ showing multistability for four different initial conditions, at $\alpha=0.03$ and $w=1.5$, $\gamma=2$, chain
length $N=200$. A transient of $10^4$ time units has been discarded.
}
\label{fig:figure6}
\end{figure}

The main phenomenology is exemplified by Figures \ref{fig:figure1}, \ref{fig:figure2} and  \ref{fig:figure3} for a particular 
realization of the disorder. The first observation 
is that, for relatively large $\gamma$, the system approach a fully synchronized regime 
whereby all the oscillators have the same common frequency but different
amplitudes $a_n$ and constant phase shifts $\varphi_n$. Such a state 
depends on the specific disorder realization and 
should correspond to a stable stationary solution of 
eqs.~(\ref{ampphas}) (a uniformly rotating solution of  eqs. (\ref{cgl})).

Upon decreasing $\gamma$ the global-phase locked state destabilizes
gradually. There is the 
spontaneous creation of a ``defect'' at some 
random location, where the amplitudes 
at a few neighboring sites start to oscillate 
in time (see the case $\gamma=0.74$ in figure \ref{fig:figure1},
where the position of the defect is around $n=170$). 
Further decrease of $\gamma$ leads to appearance of new defects and to more complex
dynamics at each defect  (figure \ref{fig:figure2}). As a result, a regime appears 
where patches of phase-entrained 
oscillators (frequency plateaus), are separated by defects that may oscillate
periodically or chaotically (see the labeled arrows 
in figure \ref{fig:figure1}). 

An heuristic explanation of the existence of the localized instability 
can be given by the following argument. Let denote 
by $\psi_n=u_n \exp(-i\mu t)$ the global phase-entrained
solution (existing at large values of $\gamma$), 
with $\mu$ being its frequency.
The linear stability analysis is performed
by letting $\psi_n=(u_n + \chi_n)\exp(-i\mu t)$ with
$\chi_n$ complex, and linearizing
the equation of motion to obtain 
\begin{equation}
\dot \chi_n = [i(\omega_n-\mu)+\gamma]\chi_n 
+ic[\chi_{n+1} + \chi_{n-1}] + (i\alpha - \beta) \left(2|u_n|^2 \chi_n 
+u_n^2 \chi_n^*\right).
\label{linear}
\end{equation}
The above eigenvalue problem
contains random terms (quenched disorder). It is thus 
akin to a random matrix problem where we expect to have 
localized eigenvectors. Also, it is not self-adjoint so
we expect complex eigenvalues leading to oscillatory instability
and thus localized modulations of the solution.

The localized nature of chaotic states is illustrated in
the left panel of Fig. \ref{fig:figure5}. We show the space-time evolution of
the Lyapunov vector associated with the largest  Lyapunov
exponent, computed by evolving the linearized equation 
(same as eq. (\ref{linear}),
but on top of a chaotic solution $\psi_n(t)$).
The appearance of chaotic oscillations in usually associated
to the fact that amplitudes tend to become small (see 
figure \ref{fig:figure3} and the right panels in
fig. \ref{fig:figure5}). So it is evident that
the amplitude dynamics plays a crucial role in this 
regime.

\begin{figure}
\begin{center}
\includegraphics[width=0.65\textwidth]{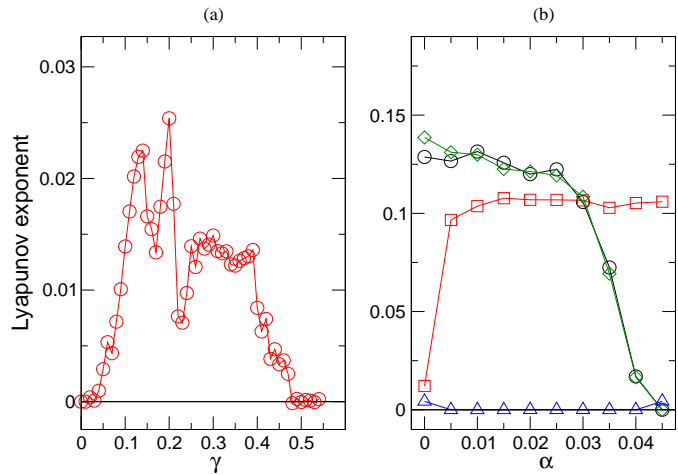}
\end{center}
\caption{(a) The largest Lyapunov exponent versus $\gamma$ for the 
same disorder realization and parameters as in previous figures.
(b) The largest Lyapunov exponent versus $\alpha$ for four different
initial conditions showing the multistable behavior $N=200, w=1.5, \gamma=2$.
The error on the exponent is roughly of the size of the symbols.
}
\label{fig:figure6b}
\end{figure}

The same type of scenario is observed 
upon changing $\alpha$ at fixed $\gamma$
(data not reported). A novel feature however emerges in the limit of 
very small $\alpha$ when the steady state depends on the initial conditions
(see figure \ref{fig:figure6}). The observed final state entails phase-entrained cluster with chaotic ones that resemble spatio-temporal intermittency. Multistable behavior is confirmed by computing the 
Lyapunov exponent $\lambda$. As seen in figure \ref{fig:figure6b},
regular trajectories with $\lambda\approx 0$ coexist with chaotic ones
for the same value of the parameter $\alpha$ suggesting coexistence
of regular and chaotic attractors.

The above results have been obtained for an uniform and thus 
bounded distribution
of frequencies.  Actually, the choice of the distribution is relevant. 
For instance, a set of simulations performed with a
Gaussian distribution with the same width $w$ indicate that it is 
somehow easier to find unlocked states there. 

\section{Ensemble properties}
\label{sec:ens}
\begin{figure}
\begin{center}
\includegraphics[width=0.8\textwidth]{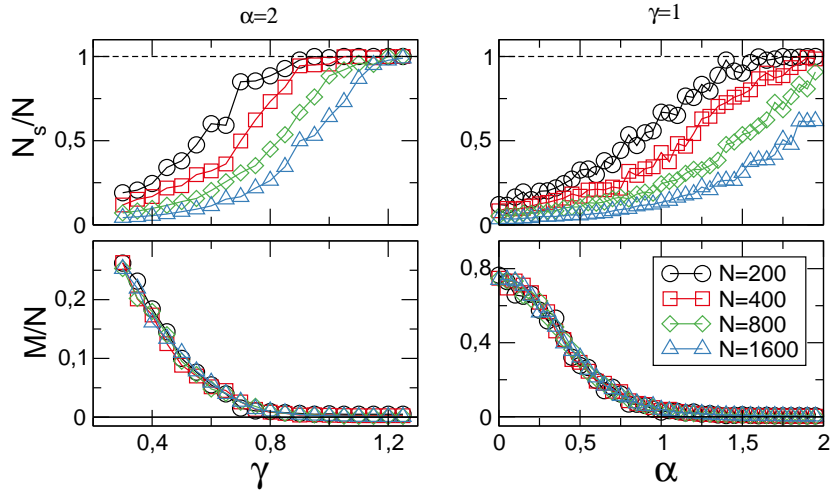}
\end{center}
\caption{Upper panels: the disorder-averaged size of the 
larger phase-locked cluster sizes; lower panels:
the average number of clusters as a function  
$\gamma$ and $\alpha$. Different symbols refer to different
lattice sizes (see legend). The threshold to locate the clusters is set to $10^{-3}$. 
}
\label{fig:figure7}
\end{figure}

In the section above we have focused on a specific realization
of the random frequencies $\omega_n$. 
For a statistical characterization, we measured the length of the 
phase-locked clusters $l_i$ defined by the condition that the mean
frequency difference is less that a given threshold. Let us denote 
by $M$ their number and by $N_s$ the size of the largest one.
In fig.~\ref{fig:figure7} we report two series of simulations 
averaged over disorder realizations. The data show 
a significant dependence on the lattice size: full phase-locking is achieved
provided that $\gamma$ or $\alpha$ are large enough but
the value required increases systematically with $N$.
On the other hand, the number of clusters $M$ seems
to be extensive in the size. This also means that the 
average density of localized (regular or chaotic) 
defects is roughly constant but vanishes for 
large  $\gamma$ or $\alpha$. 

\begin{figure}[t]
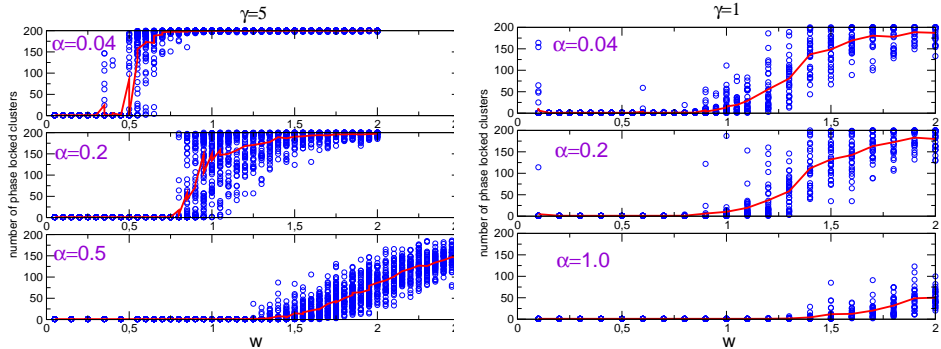

\begin{center}
\includegraphics[width=0.45\textwidth]{distrib_cluster_w_g5}
\includegraphics[width=0.45\textwidth]{distrib_cluster_w_g1}
\end{center}
\caption{The dependence of phase-locked cluster number on the disorder strength $w$
for a chain of length $N=200$, 
for two different values of $\gamma=1$ and $5$ and different $\alpha$.
Each point represents a different disorder realization, the solid lines are the 
corresponding averages. The threshold to locate the clusters is set to $10^{-3}$. 
Zero number of clusters (at small disorder) means full synchrony, the number of clusters close to $N=200$ (at large disorder) means almost complete asynchrony.
}
\label{fig:figure8}
\end{figure}

To appreciate the role of the disorder strength and fluctuations,
we report in Figure \ref{fig:figure8}
the number $M$ of phase-locked clusters as a function of $w$.
To have an idea of the sample-to-sample statistics, we draw $M$
for each disorder realization. There is a clear transition from 
full entrainment ($M=1$) upon increasing $w$ that for smaller $\alpha$
appears to be more abrupt.

\section{Phase approximation}
\label{sec:phase}
An interesting question is to what extent the dynamics can be described
in term of the phase approximation, where the amplitude evolution is neglected.
This approach allows also to rationalize the phenomenology observed 
in the above model.

For weak coupling and large enough gains, the system can indeed 
be approximated in terms of an effective equation for the relative phases 
$\varphi_n$ between neighboring oscillators.
If the amplitude relaxes quickly enough and $c$ is small, 
one can perform a straightforward adiabatic 
elimination of the amplitudes in eqs. (\ref{ampphas}):
letting first $a_n=\sqrt{\gamma/\beta}(1 + r_n)$
we approximate eqs. (\ref{ampphas}) as
\begin{eqnarray}
&& \dot r_n \approx -2\gamma r_n +c \left[\sin \varphi_n -
\sin \varphi_{n-1}  \right] \\
&& \dot\varphi_n \approx \delta_n  +2\alpha{\frac{\gamma}{\beta}} (r_{n+1}-r_{n}) +c \left[
\cos \varphi_{n+1} -
\cos  \varphi_{n-1}\right] \nonumber
\end{eqnarray}
Eliminating  the first equation adiabatically, we get 
$r_n = c \left[\sin \varphi_n -\sin \varphi_{n-1}  \right]/(2\gamma)$
and 
\begin{equation}
\dot\varphi_n =\delta_n  +{\frac{\alpha c}{\beta}} 
(\sin \varphi_{n+1} +
\sin \varphi_{n-1}- 2 \sin \varphi_{n})
+c \left[
\cos \varphi_{n+1} -
\cos  \varphi_{n-1}\right]
\label{phas}
\end{equation}
which is akin to the Kuramoto-Sagacuchi model \cite{sakaguchi1986soluble,bolotov2019dynamics},
In the case with no disorder $\delta_n=0$, this 
equation has bee studied in detail  
in ref. \cite{pikovsky2006phase}.
As discussed there, the  terms in  $\sin \varphi$ (dissipative couplings) 
favor synchronization while those in $\cos \varphi$ (dispersive or conservative couplings) hinders it. This is best understood by considering the 
``conservative limit'' where the frequency dispersion is small 
$\alpha \to 0$.  The phase equation is further approximated as  
\begin{equation}
\dot\varphi_n \approx \delta_n  +c \left[
\cos \varphi_{n+1} -
\cos  \varphi_{n-1}\right]
\end{equation}
which has vanishing phase-space divergence, meaning that the dynamics is similar to a (disordered) conservative system. The physical interpretation is that in this limit, the amplitude and the phase dynamics decouple from the beginning,
and the phase dynamics is purely conservative because the coupling is purely reactive in eq. (\ref{cgl}).  
Since the phase space is compact (variables are angles), a small dissipation
will stabilize motion on tori and we may expect some form of multistability
as in the case of Hamiltonian systems with weak dissipation \cite{Feudel96multistab}. 
This qualitatively fits with 
the phenomenology observations made in the previous sections, see
again figures \ref{fig:figure6} and \ref{fig:figure6b}.

An analysis of eq. (\ref{phas}) in presence of the random detunings 
$\delta_n$ has been given recently in ref.\cite{bolotov2019dynamics}.
It is there shown that the stationary
synchronized solution persists upon increasing 
the level of disorder in the natural frequencies. 
and that the phase shifts. Also, the domain 
of existence of the stable
synchronous regime widens in comparison to the
case of zero phase shift. This is quite counterintuitive,
because a phase shift terms usually act against
synchronization.

Note that equation (\ref{phas}) is, in this approximation, independent on $\gamma$.
We remark that in the case in which $\gamma \to \gamma_n$ is not
uniform, we expect the nearest-neighbor couplings to be site dependent,
thus representing a further source of inhomogeneity in the 
phase equations. Moreover,
as in the case of the well-know Adler equation, 
it gives a rough criterion: stationary solutions exist only if 
$\delta_n$ is not too large. A necessary condition for locking would be
$ w < 4\frac{\alpha c}{\beta}$ as the frequencies are distributed uniformly.

To conclude this Section we add some remarks on the continuum limit
of the derived phase equations.
For small $\varphi_n$ the equation is approximated
to quadratic order as
\begin{equation}
\dot\varphi_n \approx \delta_n  +{\frac{\alpha c}{\beta}} 
(\varphi_{n+1} +\varphi_{n-1}- 2  \varphi_{n})
-\frac{c}{2}(\varphi^2_{n+1} -\varphi^2_{n-1})
\end{equation}
the first term reminds of a diffusion equation with random sources. The 
nonlinear term is dispersive. Taking the continuum limit, 
letting $D={\frac{\alpha c}{\beta}}$ and denoting 
by $\Delta(x)$ the random frequency differences, we have
\begin{equation}
\dot \varphi = \Delta(x) + D\partial_x^2 \varphi - c \partial_x(\varphi^2)
\label{burgers}
\end{equation}
which is incidentally a Burgers equation with 
\textit{quenched} static disorder. As it is well know, the Burgers equation
can be mapped onto the Kardar-Parisi-Zhang (KPZ) equation 
via the transformation $\varphi=\partial_x h$ yielding
\begin{equation}
\dot h = \omega(x) + D\partial_x^2 h - c(\partial_x h)^2 + F
\end{equation}
where $F$ is an integration constant. The relation   to coupled oscillator
lattices with KPZ dynamics has been pointed out before \cite{pikovsky2003synchronization} and has been invoked to explain
phase dynamics in the noisy Kuramoto-Sakaguchi model with local couplings
\cite{lauter2017kardar}. It should be emphasized that here we are
dealing with the case of KPZ with quenched 
additive noise it usually referred to as \textit{columnar disorder}
in the literature \cite{krug1997origins}.

Finally, An alternative formulation is obtained taking the time derivative
of (\ref{burgers}) and introduce the new field $v$
\begin{eqnarray}
&&\dot \varphi = v\\
&&\dot v = D\partial_x^2 v - 2c \,\partial_x(v\varphi)
\nonumber
\end{eqnarray}
from which it is seeen that $v=0$, $\bar\varphi(x)$ is the stationary frequency-entrained solution. This introduces some form 
of inertia in the phase dynamics.  
This will not be discussed further but we remark that this equation is 
different from the one given in eqs. (36-37) of  ref. \cite{hong2005collective}.
This suggests that the model considered here may belong to a different 
class.

\section{Conclusions}
\label{sec:conclusions}

In summary, we have studied synchronization transitions in a disordered Ginzburg-Landau lattice
of coupled limit-cycle oscillators. We have demonstrated that at large values of the activity
parameter $\gamma$ a fully synchronous state in the lattice establishes. For smaller values
of $\gamma$, transition to turbulence occurs through a sequence of appearing isolated 
periodic  or chaotic ``defects''. At the intermediate stage the patches between the defects remain synchronized, so
that the dynamics can be described as a combination of several synchronous clusters having different frequencies, plus
several localized regions with weak chaos. In the latter regions the dynamics of the amplitudes appears
to be crucial. Thus, in the phase approximation, where the amplitudes are considered as slaved variables,
the features of localized chaotic defects are not properly reproduced. We have also demonstrated how
the number of different clusters in the lattice grows with the growing strength of disorder. 

\section*{Declaration of Competing Interest}
The Authors declare that they have no known competing 
financial interests or personal relationships that could have appeared to 
influence the work reported in this paper.

\section*{Acknowledgements}

SL acknowledge the research scholarship 91673361, DAAD funding programme \textit{Research stays for university academics and
scientists}, 2017 for granting the sojourn at Potsdam University
where this work was initiated. AP was supported by Russian Science Foundation (grant Nr. 17-12-01534) and by DFG
(grant PI 220/22-1).

\bibliography{sref}

\end{document}